# Structural defects and electronic structure of N-ion implanted TiO$_2$: bulk versus thin film

D.A. Zatsepin[1,2], D.W. Boukhvalov[3,4], E.Z. Kurmaev[1,2], I.S. Zhidkov[2], N.V. Gavrilov[5],

M.A. Korotin[1], S.S. Kim[6]

[1] *M.N. Miheev Institute of Metal Physics, Ural Branch of Russian Academy of Sciences, 18 Kovalevskoj Str., 620990 Yekaterinburg, Russia*
[2] *Technological Institute, Ural Federal University, Mira Street 19, 620002 Yekaterinburg, Russia*
[3] *Department of Chemistry, Hanyang University, 17 Haengdang-dong, Seongdong-gu, Seoul 133-791, Republic of Korea*
[4] *Theoretical Physics and Applied Mathematics Department, Ural Federal University, Mira Street 19, 620002 Yekaterinburg, Russia*
[5] *Institute of Electrophysics, Russian Academy of Sciences–Ural Division, 620016 Yekaterinburg, Russia*
[6] *School of Materials Science and Engineering, Inha University, Incheon 402-751, Republic of Korea*

Systematic investigation of atomic structure of N-ion implanted TiO$_2$ (thin films and bulk ceramics) was performed by XPS measurements (core levels and valence bands) and first-principles density functional theory (DFT) calculations. In bulk samples experiment and theory demonstrate anion N→O substitution. For the thin films case experiments evidence valuable contributions from N$_2$ and NO molecule-like structures and theoretical modeling reveals a possibility of formation of these species as result of the appearance of interstitial nitrogen defects on the various surfaces of TiO$_2$. Energetics of formation of oxygen vacancies and its key role for band gap reduction is also discussed.

**1. Introduction**

One of the most potential applications of TiO$_2$ is its catalytic capability of splitting water into oxygen and hydrogen [1]. However, pure TiO$_2$ can be activated only by ultraviolet light because of its large band gap (3.0 and 3.2 eV for rutile and anatase, respectively). The nitrogen doping induces the band gap narrowing, and shifts the photocatalytic response of TiO$_2$ towards the visible region of solar spectrum [2]. However, the doped nitrogen atoms can not only substitute a lattice oxide anion but also occupy interstitial positions forming molecular N$_2$ and NO species [3-5]. The similar structural defects were found in nitrogen doped oxides SiO$_2$ [6] and ZnO [7]. According to recent work [8] the introducing of large portion of chemisorbed N$_2$- molecules instead of atomic N-states could lead to the destruction of the crystal structure of initial $TiO_2$. Another problem for controllable synthesis of N-doped $TiO_2$ is connected with oxygen vacancies which are formed under N doping of $TiO_2$ due to replacing the oxygen by nitrogen atoms [9-11]. DFT calculations evidence that electronic structure of $TiO_2$:N strongly depends from concentration of oxygen vacancies [12]. In addition, the DFT calculations show that a significant band gap narrowing may only occur under heavy nitrogen doping [13,14]. Thus, to reduce the energy gap of the titanium dioxide which is necessary for efficient photocatalysis the following conditions should be realized: (*i*) anionic N→O substitution and minimizing the formation of $N_2$ and NO structural defects, (*ii*) the optimal concentration of oxygen vacancies and (*iii*) the high doping level.

In connection with this in the present paper we have studied the formation of structural defects and electronic structure in nitrogen ion implanted $TiO_2$ (both in thin film and bulk ceramic) under different fluencies with help of X-ray photoelectron spectroscopy (core levels and valence bands). The obtained results are compared with DFT calculations of formation energies and band gap values for different configurations of structural defects and concentration of oxygen vacancies.

**2. Experimental and Calculated Details**

$TiO_2$ bulk ceramic samples were prepared by hot pulsed pressing of appropriate ceramic powders obtained by electrical explosion of wires in the oxygen-containing media. After processing the plates were annealed within 1 hr at 1040 °C. The samples were on average 13 mm in diameter and 1–2 mm in height; their density was 4.25 g cm$^{-3}$. We had verified phase composition by X-ray diffraction (XRD) technique, and the final $TiO_2$ host was found to be nearly all single-phase rutile (99.85%). The parameters of tetragonal lattice were determined as $a = 4.592$ Å and $c = 2.960$ Å. Finally, the average crystallite size was found to be >200 nm.

TiO$_2$ coatings were made with the help of a sol–gel chemical technique in which titanium isopropoxide (97%), nitric acid (60%), and anhydrous ethanol had been used as the precursor, catalyst, and solvent, respectively. The purified and deionized water was used to hydrolyze the precursor mentioned, and all chemicals were used as received without any further purification. So the TiO$_2$ films were deposited on Si-wafers (100) by means of a dip coating process, and 1-butanol was added to the coating sols in order to control their wettability and the viscosity. The substrates were ultrasonically cleaned within 30 min in acetone and ethanol in sequence. After that they were then washed with deionized water. The withdrawal rate of the substrate was nearly 4 mm s$^{-1}$. Finally, the as-prepared films were dried at room temperature and kept in an oven at 60 °C for 1 d to remove the remaining solvents completely; they were then annealed at 100 °C for 2 h. The anatase films were characterized by field emission scanning electron microscopy and atomic force microscopy to confirm that high-quality films were produced. Refer to Ref. 9 for full details on TiO$_2$ film synthesis and characterization.

X-ray photoelectron spectroscopy (XPS) measurements were performed using a PHI XPS Versaprobe 500 spectrometer (ULVAC-Physical Electronics, USA) with a quartz monochromator and analyzer guaranteed working in the range of binding energies from 0 to 1500 eV with an energy resolution of $\Delta E \leq 0.5$ eV for Al $K\alpha$ radiation (1486.6 eV) and a vacuum of 10$^{-7}$ Pa. The Al $K\alpha$ X-ray spot size was 100 μm in diameter with an X-ray power load delivered to the sample not more than 25 W (the mode have been calibrated by manufacturer, ULVAC-Physical Electronics, USA). Thus the typical XPS signal-to-noise ratios were not worth than 11000 : 4. The residual XPS background (BG) was removed using the Tougard approach with Doniach-Sunjic line-shape asymmetric admixture (well described elsewhere). After that, the XPS spectra were calibrated using reference energy of 285.0 eV for the carbon 1$s$ core-level.

We performed the DFT-calculations by employing the SIESTA pseudopotential code, well described in Ref. 15, because this technique has been successfully applied in the previous studies of doped semiconductors [16]. We used the Perdew–Burke–Ernzerhof variant of the generalized gradient approximation (GGA-PBE) [17] for the exchange-correlation potential with a full optimization of the atomic positions. During our calculations, the electronic ground state was consistently found using norm-conserving pseudopotentials for the cores and a double-ξ plus polarization basis for localized orbitals for Ti, N and O. The appropriate parameters (forces and total energies) were optimized with an accuracy of 0.04 eV Å$^{-1}$ and 1.0 meV, respectively. For atomic structure calculations, we employed pseudopotentials for O, N and Ti. The formation energies ($E_{form}$) were determined by considering the supercell both "with" and "without" a given defect. For realistic calculations we had employed the $Ti_{32}O_{64}$ with a rutile structure for simulating the "bulk" phase and "slabs" of few $TiO_2$ layers. The same number of atoms in anatase phase for the modeling of (100), (010) and (001) surfaces was taken (see Fig. 1). Based on our previous findings [16], the various combinations of structural defects (single substitutional (S) and interstitial impurities (I) as well as and their combinations) were calculated, taking into account the presence of the oxygen vacancies (vO).

The CPA-calculations [18] were performed for stoichiometric $TiO_2$ and nitrogen doped $TiO_{2-y-\delta}N_y$ (y(δ)-0.03; 0.06) with rutile structure with help of an in-house code developed in Ref. 18. We used the values of the radii defined by default for the muffin-tin (MT) spheres $R_{Ti}$=2.42 au and $R_O$=1.85 au and also two types empty spheres (ES) $R_{ES}$=1.89 and 1.71 au. In the calculation the Ti (4s, 4p, 3d)-states, O (2s, 2p, 3d) and ES (1s, 2p, 3d)-states were included as the basis functions. We have calculated band structure for undoped $TiO_2$ and two N-doped compounds: $TiO_{1.88}N_{0.06}vO_{0.06}$ and $TiO_{1.91}N_{0.06}vO_{0.03}$ with different ratio of nitrogen content and oxygen vacancies denoted in chemical formulas as vO.

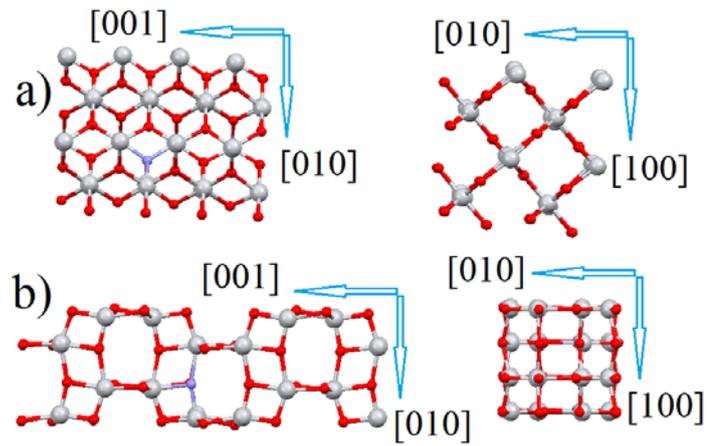

**Figure 1.** (a) Views alone different axes of supercells for bulk rutile $TiO_2$; (b) The (001) surface of anatase $TiO_2$ with a single nitrogen impurity.

## 3. Results and Discussion

In Figure 2 the XPS survey spectra of N-ion implanted $TiO_2$ are presented. As seen, besides oxygen and titanium lines the N 1s-signal is revealed which evidences about incorporation N-atoms to bulk and thin film of ion-implanted $TiO_2$ and no visible contributions from other impurities were found. The surface composition determined from these spectra is shown in the Table 1.

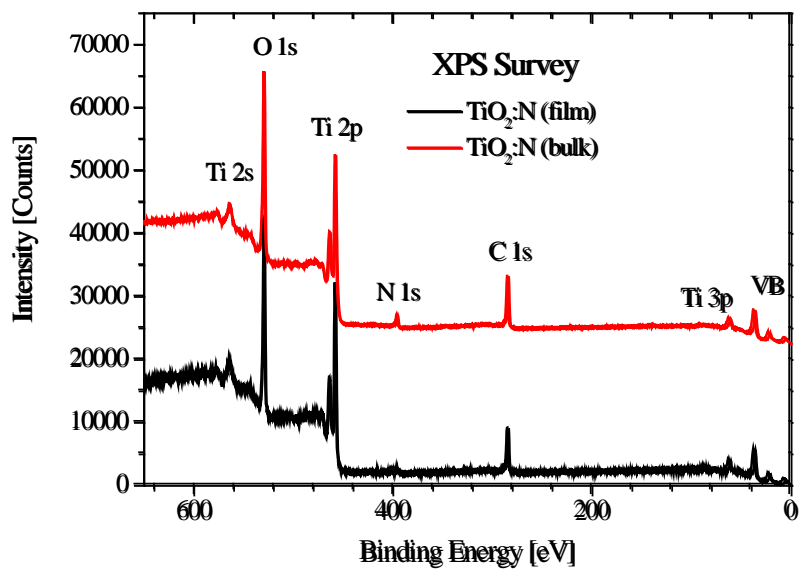

**Figure 2.** XPS survey spectra of nitrogen ion implanted bulk and thin film of $TiO_2$.

**Table 1.** Surface composition of nitrogen implanted $TiO_2$ thin film and ceramics (at.%).

| Sample | Ti | O | N | C |
|---|---|---|---|---|
| $TiO_2$:N (film), D=$1\cdot10^{17}$ cm$^{-2}$ | 13.6 | 55.3 | 1.6 | 29.5 |
| $TiO_2$:N (bulk), D=$1\cdot10^{18}$ cm$^{-2}$ | 20.4 | 40.8 | 5.4 | 33.3 |

XPS N 1s core level spectrum of ion implanted $TiO_2$ thin film (Fig. 3a) has a rather complicated structure with well pronounced three peaks located at 396.3, 399.8 and 403.0 eV. The peak centered at 396.3 eV is close to that of TiN [18] and therefore can be attributed to substituted nitrogen atoms which form N-Ti bonds in ion implanted $TiO_2$. According to Refs. 4 and 13, the ~400 eV peak could be result of bonding of interstitial N-atoms to O sites. *Nambu et al.* [5] identified ~403 eV peak to molecular nitrogen $N_2$ by means of near edge X-ray absorption spectroscopy (NEXAFS). One can see from Fig. 3a a strong reduction of intensity of the $N_2$ and NO related peaks for ceramic sample with respect to those of thin film.

Figure 3b shows XPS Ti 2p core level spectra for undoped and N-ion implanted $TiO_2$. The energy position of the main $2p_{3/2}$ and $2p_{1/2}$ peaks are identical to those of undoped $TiO_2$ with $Ti^{4+}$-states. For ion implanted samples one can see the appearance of high-energy shoulder at 456.7 eV which can be attributed to $Ti^{3+}$-states [14].

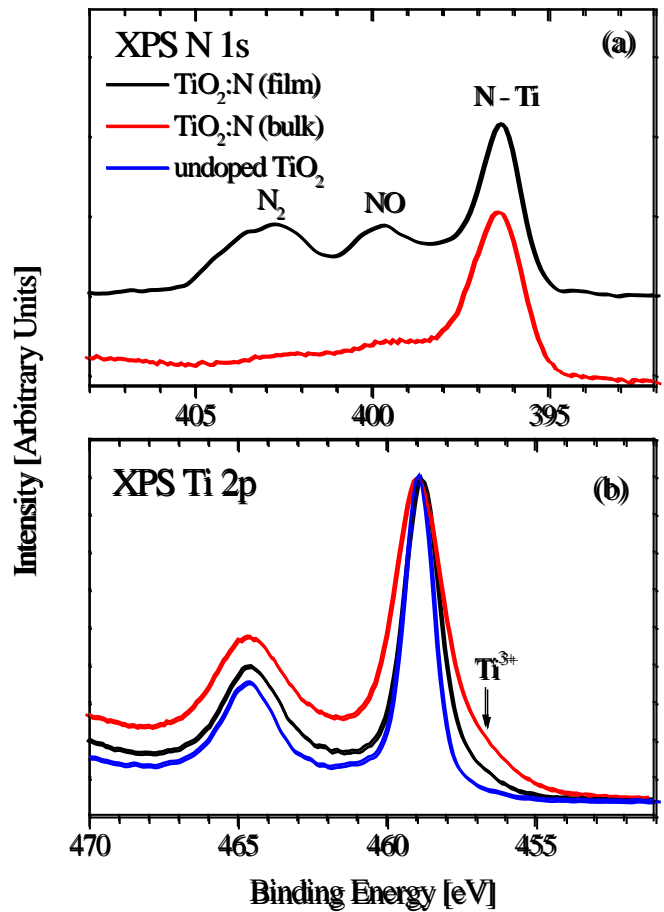

**Figure 3.** XPS N 1s (upper panel) and Ti 2p (lower panel) spectra of undoped and nitrogen ion implanted $TiO_2$.

XPS valence band spectrum of nitrogen ion implanted $TiO_2$ thin film is found to be very similar to that of undoped $TiO_2$ (Fig. 4). On the other hand, XPS VB of ion-implanted ceramic sample prepared at higher fluence shows the presence of additional features: the subbands located at ~16 and 0.5 eV. The energy position of these features is close to those of $TiN_x$ [19] and can be attributed to N 2s and Ti 3d-subbands.

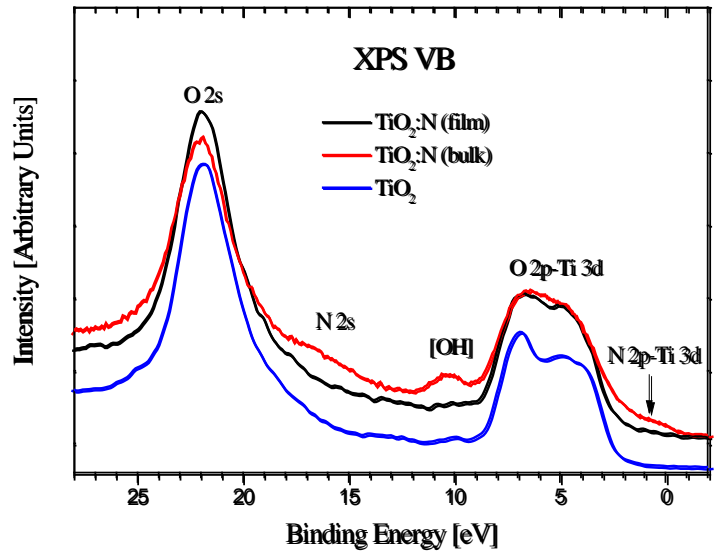

**Figure 4.** X-ray photoelectron spectra of valence bands (XPS VB) for N-ion implanted $TiO_2$.

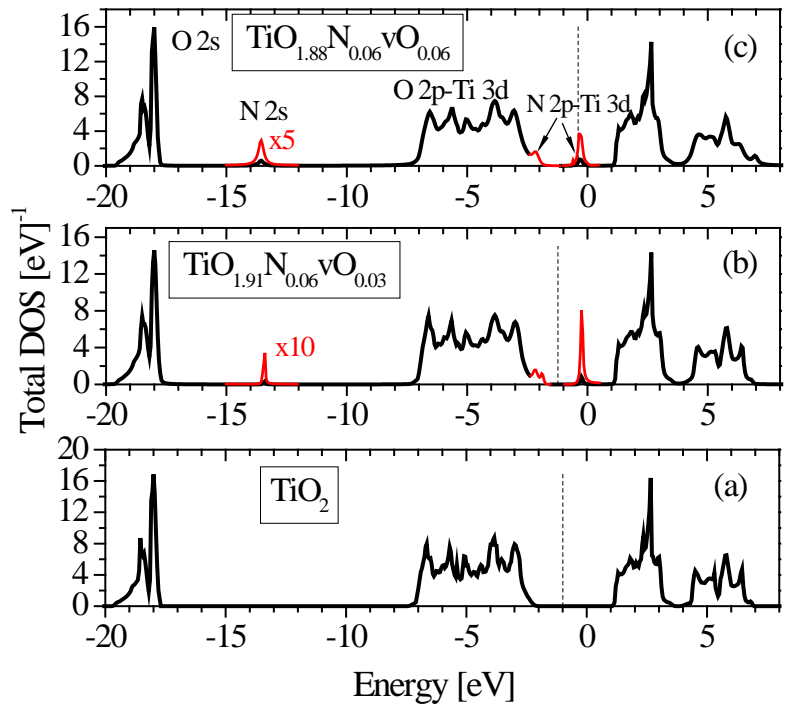

**Figure 5.** The calculated total densities of states (DOS) for undoped and nitrogen doped bulk $TiO_2$ (dashed line corresponds to the Fermi level).

The main experimental features of XPS VB of pure $TiO_2$ are well reproduced by CPA calculation (Fig. 5a). The appearance of additional N 2s-states in between of O 2s-states and bottom of the valence band and hybridized N 2p-Ti 3d-O 2p states at the top of the valence band in calculated total DOS of N-doped $TiO_2$ are found for both $TiO_{1.88}N_{0.06}vO_{0.06}$ and $TiO_{1.91}N_{0.06}vO_{0.03}$ compositions. However for $TiO_{1.88}N_{0.06}vO_{0.06}$ with equal ratio of nitrogen and oxygen vacancies the Fermi level crosses the impurity N 2p-Ti 3d subband (Fig. 5c) which corresponds to metallic bonding. On the other hand, for $TiO_{1.91}N_{0.06}vO_{0.03}$ composition with ratio of nitrogen atoms and oxygen vacancies 2:1 (Fig. 5b) the Fermi level is located within gap between hybridized N 2p-Ti 3d-O 2p states and N 2p-Ti 3d impurity states reducing the band gap value of pristine $TiO_2$ (Fig. 5a). Therefore the ratio of nitrogen atoms and oxygen vacancies is found to be critical for band gap reduction of bulk $TiO_2$;N.

However, in CPA calculations we could simulate only substitutional effect in $TiO_2$:N when doped nitrogen atom substitute oxygen atoms (N→O) with formation of oxygen vacancies which according to XPS N 1s spectra (Fig. 3) takes place under doping of bulk $TiO_2$. On the other hand, XPS N 1s spectra of $TiO_2$:N thin films definitely show the presence of additional features which can be attributed to embedding of N-atoms to interstitials and formation of $N_2$ and NO species. In connection with this we have calculated the formation energies of different nitrogen defects both for bulk and surface of $TiO_2$ and then basing on the obtained results have performed DFT calculations taking into account the real configurations of structural defects.

**Table 2.** Formation energies (eV/defect) for various combinations of defects in bulk and surface of $TiO_2$. Substitutions are denoted as S(substituted atom), interstitial as I, and oxygen vacancies as vO. The most probable configurations of defects for each type of hosts are marked by bold.

| Defect    | Bulk rutile | Anatase surface (100) | Anatase surface (001) |
|-----------|-------------|-----------------------|-----------------------|
| S(O)      | 4.43        | 4.90                  | 4.50                  |
| S(O)+vO   | **3.35**    | 3.20                  | 3.36                  |
| S(Ti)     | 6.44        | 6.63                  | 5.26                  |
| S(O)+I    | 3.91        | **2.01**              | **1.17**              |
| S(O)+I+vO | 3.44        | 2.58                  | 1.78                  |

The first step of our DFT modeling was a search of most probable configurations of structural defects in bulk rutile. The results of calculations (Table 2) demonstrate that the presence of oxygen vacancies always provide the decreasing of formation energies whereas the «pure» substitution of titanium atom by nitrogen S(O) is energetically unfavorable. For placement of nitrogen atoms in the interstitial void the combination of removed oxygen atom and interstitial nitrogen atom S(O)+I+vO is more favorable than S(O)+I. Thus we can summarize that for the case of bulk rutile the implantation of nitrogen will provide the various combinations of N→O substitutions with oxygen vacancies.

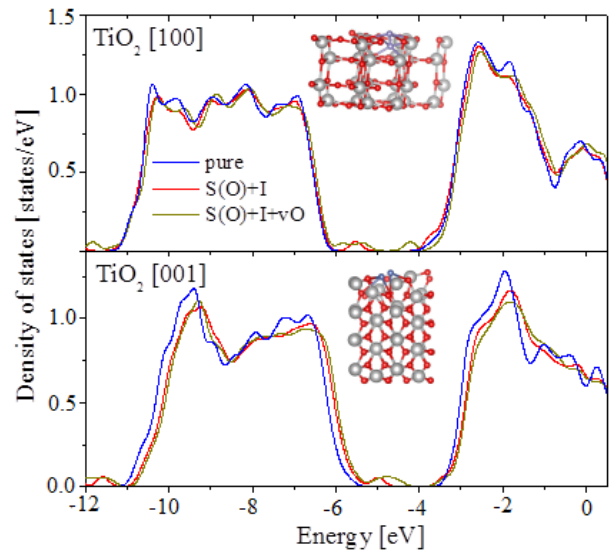

**Figure 6.** Total densities of states (DOS) of $TiO_2$ surfaces pure and with implanted pair of substitutional and interstitial nitrogen atoms (see insets) with and without oxygen vacancy in vicinity.

The next step of our modeling is the check the effect of implantation of nitrogen on the surface of anatase phase by calculations of the energetics of the same configurations of defects on the surface and subsurface layers. The energetics of defects on (100) and (010) surfaces is almost the same and further we will discuss only (100) and (001) surfaces. Similar to the case of transitional metals impurities in $TiO_2$ bulk and thin film [15] the formation energies for nitrogen implanted to (100) and (001) anatase surface are found to be lower than in the case of bulk rutile (Table 2). The single interstitial nitrogen impurities form bonds not only with titanium but also with oxygen atoms with bond length close to that of NO molecule and pair of substitutional and interstitial nitrogen impurities – the $N_2$-like pair (see inset on Fig.6). These results can explain an appearance of the corresponding peaks in XPS N 1s spectra in $TiO_2$:N thin films (Fig. 3). Unlike to the case of bulk (Fig. 5) the structural defects on the surface provide the reasonable decreasing of band gap only for S(O)+I configuration (see Table 3 and Fig. 6). The additional formation of oxygen vacancies S(O)+I+vO does not provide the further decreasing of formation energies (see Table 2) and proves the band gap narrowing (Fig. 6) from 1.22 eV to 0.67 eV for (100) surface and from 1.15 eV to 0.65 eV for (001) surface. Usually the DFT calculations underestimate the band gap values [19] but we can determine the relative values of band gap reduction which is found to be ~1.15-1.22 eV for S(O)+I configuration (close to experimental value of 0.8-0.9 eV [20]) and ~0.65-0.67 eV for S(O)+I+vO configuration which certainly overestimates the band gap narrowing. Therefore in the case of surface the propagated defects will be formed without oxygen vacancies that are in a qualitative agreement with smaller decrease of band gap values.

Table 3. The calculated band gap values (in eV) for (100) and (001) surfaces of anatase.

| Configuration | (100) | | | (001) | | |
|---|---|---|---|---|---|---|
| | VB | CB | ΔE | VB | CB | ΔE |
| Pure | -6.11 | -3.98 | **2.13** | -5.44 | -3.52 | **1.92** |
| S(O)+I | -5.26 | -4.04 | **1.22** | -4.72 | -3.57 | **1.15** |
| S(O)+I+vO | -5.06 | -4.39 | **0.67** | -4.27 | -3.62 | **0.65** |

**4. Conclusions**

Based on combination of experimental measurements and theoretical calculations we demonstrate that implantation of nitrogen in bulk $TiO_2$ provides mainly N→O substitution. The calculations show that combination of nitrogen defects with oxygen vacancies provides the small formation energies and induces the decreasing of band gap. In contrast to bulk TiO$_2$:N the N-ion implantation of thin film causes in addition to N→O substitution the embedding of nitrogen impurities in interstitials which provides the formation of molecule-like NO and N$_2$ species detected in XPS N 1s spectra. In contrast to the case of bulk the formation of oxygen vacancies in the vicinity of embedded nitrogen atoms in thin film is less energetically favorable and overestimates the band gap narrowing. Thus implantation of nitrogen in bulk TiO$_2$ provides mainly N→O substitution, minimizes the embedding of impurity N-atoms in interstitials and gives an optimal decrease of band gap which makes this material attractive for photocatalysis and solar cells. On the other hand the interstitial nitrogen impurities and N$_2$-pairs on the surface of TiO$_2$ films could form active centers which are important for catalysis.

**Acknowledgements**

This study was supported by a grant of the Russian Science Foundation (Project No. 14-22-00004) and Russian Federation Ministry of Science and Education (Government Task 3.2016.2014/K).

**References:**

1. A. Fujishima, K. Honda, Nature Electrochemical Photolysis of Water at a Semiconductor Electrode 238 (1972) 37-38.
2. R. Asahi, T. Morikawa, T. Ohwaki, K. Aoki, Y. Taga, Science Visible-Light Photocatalysis in Nitrogen-Doped Titanium Oxides 293 (2001) 269271.
3. C. Bittencourt, M. Rutar, P. Umek, A. Mrzel, K. Vozel, D. Arcon, K. Henzler, P. Krüger and P. Guttmann, RSC Advances Molecular nitrogen in N-doped TiO$_2$ nanoribbons 5 (2015) 23350-23356.
4. S. Livraghi, M.C. Paganini, M. Chiesa and E. Giamello, Trapped molecular species in N-doped TiO$_2$ Res. Chem. Intermed. 33 (2007) 739-747.
5. A. Nambu, J. Graciani, J. A. Rodriguez, Q. Wu, and E. Fujita N doping of TiO2(110): photoemission and density-functional studies J. Chem. Phys. 125 (2006) 094706.
6. Y. Chung, J. C. Lee, H. J. Shin, Appl. Phys. Lett. Direct observation of interstitial molecular N$_2$ in Si oxynitrides 86 (2005) 022901.
7. P. Hoffmann and C. Pettenkofer Chemical nature of N-ions incorporated into epitaxial ZnO films Phys. Status Solidi B 248 (2011) 327-333.
8. K. S. Han, J. W. Lee, Y. M. Kang, Y. Y. Lee, J. K. Kang Nature of Atomic and Molecular Nitrogen Configurations in TiO$_{2–x}$N$_x$Nanotubes and Tailored Energy-Storage Performance on Selective Doping of Atomic N States Small 4 (2008) 1682-1686.

9. Abdul K. Rumaiz, J. C. Woicik, E. Cockayne, H. Y. Lin, G. H. Jaffari, and S. I. Shah Oxygen vacancies in N doped anatase TiO2: Experiment and first-principles calculations Appl. Phys. Lett. 95 (2009) 262111.

10. M. A. Korotin and V. M. Zainullina Investigation of the Influence of Nonstoichiometry and Doping with Carbon and Nitrogen on the Electronic Spectrum of Rutile by the Coherent Potential Method Phys. Solid State 55 (2013) 952-959.

11. H.-C. Wu, Y.-S. Lin, and S.-W. Lin, Calculations Mechanisms of Visible Light Photocatalysis in N-Doped Anatase $TiO_2$ with Oxygen Vacancies from GGA+U Calculations Int. J. Photoenergy (2013) 289328.

12. F. Esaka, K. Fuuya, H. Shimada, M. Imamura, N. Matsubayashi, H. Sato, A. Nishijima, A. Kawana, H. Ichimura, and T. Kikuchi, J. Vac. Sci. Technol. A Comparison of surface oxidation of titanium nitride and chromium nitride films studied by x-ray absorption and photoelectron spectroscopy 15 (1997) 2521.

13. J. A. Rodriguez, T. Jirsak, G. Liu, J. Hrbek, J. Dvorak, and A. Maiti Chemistry of $NO_2$ on Oxide Surfaces: Formation of $NO_3$ on $TiO_2$(110) and $NO_2$↔O Vacancy Interactions J. Am. Chem. Soc. 123 (2001) 9597.

14. A. Petala, D. Tsikritzis, M. Kollia, S. Ladas, S. Kennou, D. I. Kondarides Synthesis and characterization of N-doped $TiO_2$ photocatalysts with tunable response to solar radiation Original Appl. Surf. Sci. 305 (2014) 281-291.

15. J. M. Soler, E. Artacho, J. D. Gale, A. García, J. Junquera, P. Ordejón, D. Sánchez-Portal The SIESTA method for ab initio order-N materials simulation J. Phys.: Condens. Matter. 14 (2002) 2745.

16. B. Leedahl, D.A. Zatsepin, D. W. Boukhvalov, E.Z. Kurmaev, R. J. Green, I.S. Zhidkov, S. S. Kim, L. Cui, N. V. Gavrilov, S. O. Cholakh, A. Moewes Study of the Structural Characteristics of 3d Metals Cr, Mn, Fe, Co, Ni, and Cu Implanted in ZnO and $TiO_2$—Experiment and Theory J. Phys. Chem. C 118 (2014) 28143-28151.

17. J. P. Perdew, K. Burke, M. Ernzerhof Generalized Gradient Approximation Made Simple Phys. Rev. Lett. 77 (1996) 3865-3871.

18. M. A. Korotin, N. A. Skorikov, V. M. Zainullina, E.Z. Kurmaev, A. V. Lukoyanov and V. I. Anisimov Electronic structure of nonstoichiometric compounds in the coherent potential approximation JETP Lett. 94 (2012) 806-810.

19. M. J. Vasile, A. B. Emerson, F. A. Baiocchi The characterization of titanium nitride by x-ray photoelectron spectroscopy and Rutherford backscattering J. Vac. Sci. Technol. A 8 (1990) 99.

20. M. van Schilfgaarde, T. Kotani, S. Faleev Quasiparticle Self-Consistent G W Theory Phys. Rev. Lett. 96 (2006) 226402.

21. N. Delegan, R. Daghrir, P. Drogui, and M. A. El Khakani Bandgap tailoring of in-situ nitrogen-doped $TiO_2$ sputtered films intended for electrophotocatalytic applications under solar light Appl. Phys. Lett. 116 (2014) 153510.